\documentclass{ws-mpla}
\usepackage[super,compress]{cite}
\usepackage[breaklinks]{hyperref}
\hypersetup{colorlinks,urlcolor=black,citecolor=black,linkcolor=black,filecolor=black}
\usepackage{breakurl}
\usepackage{url}
\usepackage{graphicx}
\usepackage{dcolumn}% Align table columns on decimal point
\usepackage{bm}% bold math
\usepackage{color}
\usepackage{float}
\usepackage{amsmath}
\usepackage{lineno}
\usepackage{bm}

\begin{document}
%\linenumbers
\markboth{Z. M. Li}
{Overview of intermittency analysis in heavy-ion collisions}

%\markboth{Authors' Names}
%{Instructions for Typing Manuscripts (Paper's Title)}

%%%%%%%%%%%%%%%%%%%%% Publisher's Area please ignore %%%%%%%%%%%%%%
\catchline{}{}{}{}{}
%%%%%%%%%%%%%%%%%%%%%%%%%%%%%%%%%%%%%%%%%%%%%%%%%%%%%%%%%%%%%%%%%%%

\title{Overview of intermittency analysis in heavy-ion collisions}

\author{\footnotesize Zhiming Li\textsuperscript{$\star$}}
\address{Key Laboratory of Quark and Lepton Physics (MOE) and Institute of Particle Physics, \\Central China Normal University, Wuhan 430079, China\\
* lizm@mail.ccnu.edu.cn}

\maketitle

\pub{Received (Day Month Year)}{Revised (Day Month Year)}

\begin{abstract}
In this paper, a search for power-law fluctuations with fractality and intermittency analysis to explore the QCD phase diagram and the critical point is summarized. Experimental data on self-similar correlations and fluctuations with respect to the size of phase space volume in various high energy heavy-ion collisions are presented, with special emphasis on background subtraction and efficiency correction of the measurement. Phenomenological modelling and theoretical work on the subject are discussed. Finally, we highlight possible directions for future research.

\end{abstract}

%\ccode{PACS Nos.: 24.60.Ky, 25.75.−q, 25.75.Nq}

\section{Introduction}	
One of the major goals in current heavy-ion collisions is to locate the critical point (CP) in the phase diagram of strongly interacting matter predicted by the Quantum Chromodynamics (QCD)~\cite{StephanovPD,adams2005experimental,conservecharge0,conservecharge1}. Lattice QCD has shown that there occurs a smooth crossover from hadronic phase to the Quark Gluon Plasma (QGP) phase at $\mu_B = 0$~\cite{Lattice, Crossover1,Crossover2}. QCD based model calculations indicate that the transition could be a first-order at large $\mu_B$~\cite{firstorder1,firstorder2}. The endpoint of the first-order phase transition to the crossover is referred to as the critical point~\cite{CEP1, CEP2,CEP3,CEP4}. However, Lattice QCD calculations at finite $\mu_B$ face numerical challenges in computing. The location of the CP is highly theoretically uncertain. Investigations from heavy-ion experiments, such as the Beam Energy Scan (BES) II program at the Relativistic Heavy Ion Collider (RHIC)~\cite{besII,besII2}, and phenomenological modeling are required.

Fluctuations of conserved quantities, which behave differently between the hadronic and the QGP phases, are normally considered to be promising signatures for the QCD phase transition~\cite{searchCEP1,searchCEP2,searchCEP3,searchCEP4,net_proton2010,net_proton2014,net_charge2014,net-kaon,phenix,STARPRLMoment,MomentNucl,conservecharge0,conservecharge1,fluctuation1}. The singularity at the CP, at which the transition is believed to be the second-order, may cause enhancement of fluctuations if fireballs created in the collision pass the CP region during the time evolution~\cite{fluctuation2,DFphaseT1,KJSunPLB2017,KJSunPLB2018}. In heavy-ion collisions, in analogy to the critical opalescence phenomenon observed in Quantum Electrodynamics~\cite{opale1}, large density fluctuation of late-stage baryon numbers is expected to develop for the system near the CP due to a rapid increase of correlation length~\cite{StephanovPD,corrLen2,AntoniouPRL}. Strong baryon density fluctuations should be observed in experiments at kinetic freeze-out if such fluctuations can survive during final-state interactions and hadronic evolution of the system.

It is suggested that the expected large baryon density fluctuations near the CP can be expressed in terms of density-density correlator of baryon numbers, which exhibiting a fractal and scale invariant behavior~\cite{AntoniouPRL,invariant2,invariant3}. These fluctuations are supposed to form a unique pattern of self-similar or intermittency behavior of finite state particles in high energy collisions~\cite{AntoniouPRL,AntoniouPRC2010,AntoniouPRC,AntoniouPRD}. Intermittency refers to the property of scale invariance, fractality and a stochastic nature of the underlying scaling law~\cite{invariant3}. It can be measured by calculation of a scaled factorial moment (SFM) at various system scales in transverse momentum space~\cite{AntoniouPRL,NA49EPJC}. The advantage of using SFM is that it can characterize the non-statistical fluctuations in momentum spectra, which is supposed to be connected with the dynamics of particle production.

The last decade has witnessed remarkably exciting experimental explorations of intermittency and fractality in high energy collisions. A systematic search for QCD critical fluctuations has been performed by NA49 and NA61/SHINE collaborations at CERN SPS with measurements of intermittency in A+A collisions~\cite{NA49PRC,NA49EPJC,NA61QM2019}. At RHIC energies, the studies on both charged particles~\cite{STARIntermittency} and strangeness~\cite{SSinStrangeness} in Au+Au collisions at BES-I energies are reported. With the LHC at CERN ushering in a new era of TeV-scale high energy physics, the CMS collaboration has measured SFM and intermittency in $pp$ collisions up to 8 TeV~\cite{CMSIntermittency}. It is expected that fractality or self-similarity, analogous to that encountered in complex non-linear systems, might open a new way leading towards deeper insight into the CP and phase transition of QCD.

At the same time, different phenomenological models have been used to study the unique behavior of SFMs and intermittency under various underlying mechanisms~\cite{CMCPLB,LiPLBUrQMD,GopeEPJA,IJMPE,AMPTnu,EposAHEP}. With a Critical Monte-Carlo (CMC) model belonging to the 3D Ising universality class, it is found that the self-similar or intermittency nature of particle correlations is closely related to the large baryon density fluctuation associated with the critical point~\cite{CMCPLB}. By including hadronic mean-field potentials~\cite{LiPLBUrQMD} or hydrodynamical descriptions~\cite{GopeEPJA} into the data sample of a transport UrQMD model, the result exhibits a clear self-similar behavior. It infers that intermittency is associated with the mechanism of evolution of the medium produced in heavy-ion collisions.

Intermittency is measured by calculations of SFM in momentum space at various scales. Whereas, the single-particle multiplicity spectra of late-stage particles in heavy-ion collisions are largely influenced by background effects~\cite{BialasNPB1986,KFAPPB1989,SamantaJPG2021,MarekNPA2021}. In experiments, insufficient detector efficiency~\cite{STARData,MomentEfficiency} will also modify the value of SFM. To understand the underlying physics of this measurement, a careful study on the non-critical effects to get a clean signal is needed. After that, we can compare results among different experiments and with those from theoretical or phenomenological predictions.

This paper contains a review of the present study on intermittency in heavy-ion collisions over the last years. In Section 2, we introduce the method of intermittency analysis and the necessary formalism. Section 3 discusses the estimation and subtraction of background contributions in the calculation of SFMs. In section 4, the efficiency correction formula on SFM is derived. Section 5 gives an overview of experimental measurements on intermittency in current heavy-ion collisions. Section 6 is devoted to phenomenological model results on the subject and the search for power-laws. Conclusions and outlooks are summarized in Section 7.

\section{Method of Analysis}
In high-energy collisions, a power-law density fluctuation is proposed to be detectable in momentum space through intermittency analysis. The observables of our interests in this analysis are chosen to be sensitive to the power-law singularity of the density-density correlation function. The SFM of multiplicity distributions of final state particles is suggested as one of the most suitable quantities~\cite{AntoniouPRL,NA49EPJC}. For this purpose, a selected D-dimensional momentum space is partitioned into equal-sized cells. The $q$th-order SFM is defined as:
\begin{equation}
F_{q}(M)=\frac{\langle\frac{1}{M^{D}}\sum_{i=1}^{M^{D}}n_{i}(n_{i}-1)\cdots(n_{i}-q+1)\rangle}{\langle\frac{1}{M^{D}}\sum_{i=1}^{M^{D}}n_{i}\rangle^{q}},
 \label{Eq:FM}
\end{equation}
\noindent where $M$ is the number of cells in one dimension, $n_{i}$ is the measured number of particles in the $i$th cell, and the angular bracket denotes an average over the event sample.

For the system near the critical point, it is expected to observe a scaling or power-law behavior of the SFM on the partitioned number of cells $M^{D}$, given $M$ is large enough:
\begin{equation}
F_{q}(M)\sim (M^{D})^{\phi_{q}}, M\rightarrow\infty.
 \label{Eq:PowerLaw}
\end{equation}

\noindent Here $\phi_{q}$ is called intermittency index which specifies the strength of the self-similar property. The method of using SFM to search for the critical fluctuation was firstly proposed in Refs~\cite{BialasNPB1986,BialasNPB1988} several years ago. By using the effective action of a 3D Ising system belonging to the same universality class of the QCD critical point~\cite{Gavin1993PRD,Halasz1998PRD,Karsch2001PLB}, the critical $\phi_{2}$  is predicted to be $\frac{5}{6}$ for baryon density~\cite{AntoniouPRL} and $\frac{2}{3}$ for pion density~\cite{NGNPA2005}.

Besides the power-law behavior of SFMs on the partitioned numbers as described above, another promising relation between SFMs of different orders is suggested~\cite{GLPRL,GLPRD,GLPRC}:
\begin{equation}
F_{q}(M)\propto F_{2}(M)^{\beta q}, 
 \label{Eq:FqF2scaing}
\end{equation}
where $\beta_{q}=\phi_{q}/\phi_{2}$. According to Ginzburg-Landau (GL) theory, the power-law behavior of $F_{q}(M)\sim M^{D}$ might not be observed near the QCD critical point since $\phi_{q}$ depends on particular critical parameters which would vary with temperatures of the system and are unknown in nuclear collisions~\cite{GLPRL,GLPRD}. However, $\beta_{q}$ is independent of these parameters and thus the power-law behavior of $F_{q}(M)$ on $F_{2}(M)$ is feasible to measure in experiments of nuclear-nuclear collisions.   

The scaling exponent $\nu$ specifies the power-law behavior of $F_{q}(M)$ on $F_{2}(M)$ and quantitatively describes the values of $\beta_{q}$~\cite{GLPRL,GLPRD,GLPRC}:
\begin{equation}
\beta_{q} \propto (q-1)^{\nu}. 
 \label{Eq:nuscaing}
\end{equation}

This critical exponent $\nu$, which is supposed to be independent on the chosen values of specific critical parameters, can be used to investigate the presence of intermittency in the QCD phase transition~\cite{GLPRL,GLPRD, GLPRC, nuAPLB}. Theoretical prediction for the critical value of $\nu$ is equal to 1.304 in the full phase space based on Ginzburg-Landau (GL) theory~\cite{GLPRL}, and 1.0 from calculations of a two-dimensional Ising model~\cite{GLPRD}. 

\section{Background Subtraction}
In the calculations of SFMs, it is important to estimate and subtract trivial contributions from background. It has been found that the values of the measured SFMs can be significantly modified by adding uncorrelated particles as a background to the event samples containing self-similar signals.~\cite{SamantaJPG2021}. These background effects should be eliminated in the intermittency analysis.

\subsection{Mixed Event Method}
By analyzing the NA49 and NA61 results, it is shown that the experimentally measured scaling behavior of SFMs and the intermittency index can be reproduced by adding more than $90\%$ uncorrelated random tracks into the event sample generated by the CMC model~\cite{NA49EPJC,NGNPA2020}.  In this purpose, they propose to use the mixed event method to estimate background contributions~\cite{NA49EPJC,NA61universe}. Mixed events are constructed by randomly selecting particles from original events while reproducing the same multiplicity distributions. The correlations between pairs of particles which exist in the original event, are eliminated in the mixed event samples since each particle now is chosen from different events. By assuming that multiplicity distributions of measured particles in each partitioned cell can be simply divided into background and critical contributions, the correlator $\Delta F_{q}(M)$, supposed to contain only critical contributions, is defined as:
\begin{equation}
 \Delta F_{q}(M)=F_{q}^{data}(M)-F_{q}^{mix}(M).
 \label{Eq:DeltaFq}
\end{equation}

After subtracting background contributions, the intermittency index $\phi_{q}$ can be obtained from $\Delta F_{q}(M)$ instead of $F_{q}(M)$ by using Eq.~\eqref{Eq:PowerLaw}.

\subsection{Cumulative Variable Method}
The cumulative variable method was proposed for intermittency analysis in Refs~\cite{Bialas1990CFM,Ochs1990CFM} several years ago. It has been proved to be able to effectively reduce the distortions of a non-uniform single-particle spectrum~\cite{Bialas1990CFM,RefEfficiency}.

Assuming the variable $x$, e.g., momentum $p_{x}$ or $p_{y}$, is measured by a density distribution function $\rho(x)$. The cumulative variable $X(x)$ is defined as the transformation of ~\cite{Bialas1990CFM,Ochs1990CFM}
\begin{equation}
  \large X(x)=\frac{\int_{x_{min}}^{x} \rho(x)dx}{\int_{x_{min}}^{x_{max}}\rho(x)dx},
 \label{Eq:cvariable}
\end{equation}

\noindent where $x_{min}$ and $x_{max}$ are minimum and maximum values of the selected variable $x$.

The advantage of the cumulative variable $X(x)$ is that the value of the new variable does not depend on particular variable $x$ but on its density distribution function $\rho(x)$, which provides a way to compare results among different experiments. The other benefit is that the probability distribution of $X(x)$ is uniform which can remove the dependence of intermittency on density distribution of $\rho(x)$~\cite{Bialas1990CFM}.

 In the calculation of $F_{q}(M)$ of cumulative variable in a 2D momentum region, the transverse momentum space of $p_{x}p_{y}$ is transfer to the one of $p_{X}p_{Y}$. To be clear, we denote the $F_{q}(M)$ after cumulative transformation as $CF_{q}(M)$. The process of fitting intermittency index $\phi_{q}^{c}$ from $CF_{q}(M)$ is the same as $\phi_{2}$ from $F_{2}(M)$ by Eq.~\eqref{Eq:PowerLaw}.

%%%%%%%%%%%%%%%%%Fig cumulative variable
\begin{figure}[htp]
     \centering
     \includegraphics[scale=0.6]{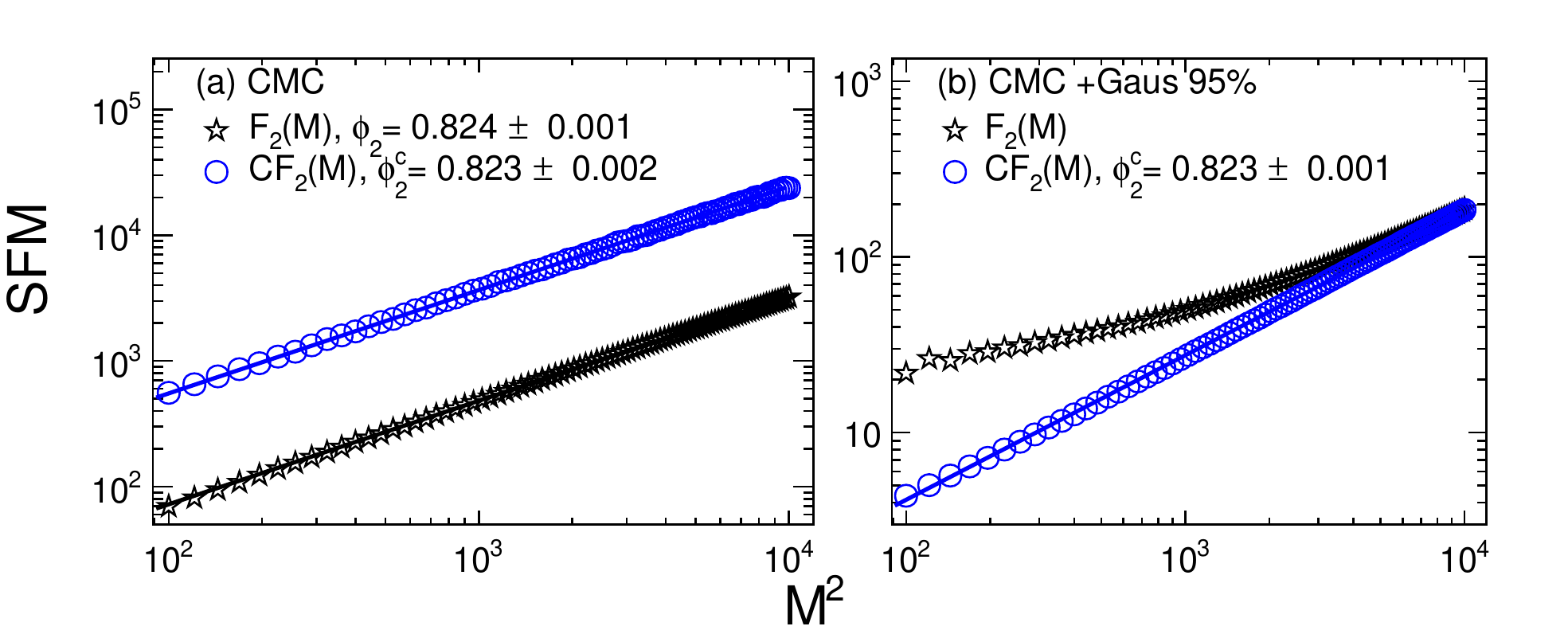}
     \caption{(a) The second-order $F_{2}(M)$ (black stars) and $CF_{2}(M)$ (blue circles) as a function of number of partitioned bins in the CMC model. (b) $F_{2}(M)$ and $CF_{2}(M)$ for the same CMC event samples but being contaminated by uncorrelated particles with a statistical Gaussian distribution.}
     \label{Fig:GasandCMC}
\end{figure}

To verify the validity of the this method, the CMC model~\cite{AntoniouPRL,CMCPLB} belonging to the 3D Ising universality is applied to generate event samples incorporating critically correlated particles in momentum space. The CMC model involves large density fluctuations risen from the self-similar correlations among particles and leads to the intermittency index of $\phi_{2}=\frac{5}{6}$~\cite{AntoniouPRL,CMCPLB}. Figure~\ref{Fig:GasandCMC} (a) shows the $F_{2}(M)$ and $CF_{2}(M)$ as a function of $M^{2}$ from the CMC model in a double-logarithmic scale. It is clearly observed that both $F_{2}(M)$ (black stars) and $CF_{2}(M)$ (blue circles) obey a good power-law dependence on the number of division bins. The fitted $\phi_{2}$ is equal to $\phi_{2}^{c}$ within statistical uncertainty. It implies that the intermittency behavior keeps unchanged by applying the cumulative variable method to a critical event sample with pure self-similar fluctuations.

We further test the method by adding background fluctuations into the signal. The CMC samples are contaminated by particles with a Gaussian distribution as background by a mixed ratio of $\lambda=95\%$. In Fig.~\ref{Fig:GasandCMC} (b), one observes that the directly calculated $F_{2}(M)$ deviates substantially from the power-law dependence on $M^{2}$ and can not be fitted by the relation of Eq.~\eqref{Eq:PowerLaw}. However, $CF_{2}(M)$ still follows the same scaling behavior with increasing $M^{2}$ as that in Fig.~\ref{Fig:GasandCMC} (a). The fitted $\phi_{2}^{c}$ is found to be equal to the one in the pure CMC event sample. It confirms that the cumulative variable method is able to effectively erase background effects in the intermittency analysis. 

\section{Efficiency Correction}
Besides background subtraction, efficiency correction for SFMs is also an important aspect in the calculations of intermittency. In experimental measurement, some particles are missing due to a limited capacity of the detector, which leads to the measured multiplicity distribution to be different from the originally produced one in the momentum space. Therefore, the calculated SFM will be changed accordingly. To recover the true SFM from the experimentally measured one, a cell-by-cell efficiency correction method is suggested in Ref.~\cite{RefEfficiency}.

The probability function of $p(n)$ is related to the one of $p(N)$ with~\cite{EfficiencyKochPRC2015,EfficiencyKochNPA}:
\begin{equation}
p(n)=\sum_{N}W(n|N)p(N).
\label{Eq:W(nN)}
\end{equation}
Here $W(n|N)$ is the probability of experimentally measured $n$ particles, given $N$ originally produced particles in the event. In general, $W(n|N)$ can be approximated by a binomial distribution as~\cite{EfficiencyKochPRC2015,EfficiencyKochNPA,EfficiencyLuo}:
\begin{equation}
W(n|N)=B(n,N;\epsilon)=\frac{N!}{n!(N-n)!}\epsilon^{n}(1-\epsilon)^{N-n},
\label{Eq:binomial}
\end{equation}
\noindent where $\epsilon$ is the particle detection efficiency.

Based on Eq.~\eqref{Eq:W(nN)} and ~\eqref{Eq:binomial}, the true factorial moment $f_{q}^{true}=\langle N(N-1)...(N-q+1)\rangle$ will be restored by dividing the measured moment $f_{q}^{measured}=\langle n(n-1)...(n-q+1)\rangle$, with certain power of the experimental detector efficiency $\epsilon$~\cite{EfficiencyLuo, EfficiencyKochPRC2015, momentHADES}: 
\begin{equation}
 f_{q}^{corrected}=\frac{f_{q}^{measured}}{\epsilon^{q}}=\frac{\langle n(n-1)...(n-q+1)\rangle}{\epsilon^{q}}.
 \label{Eq:f2correction}
\end{equation}

\noindent This method has been extensively used for efficiency corrections on the high-moment studies in experiments~\cite{EfficiencyLuo,MomentNucl,momentHADES,STARPRLMoment}.

We use the same strategy in the intermittency analysis. The $q$-th moment $\langle n_{i}(n_{i}-1)...(n_{i}-q+1)\rangle$ of SFM in each bin is corrected according to Eq.~\eqref{Eq:f2correction}. Then, the SFM defined in Eq.~\eqref{Eq:FM} can be corrected as:
\begin{eqnarray}
  F_{q}^{corrected}(M)=\frac{\langle\frac{1}{M^{2}}\sum_{i=1}^{M^{2}}\frac{n_{i}(n_{i}-1)\cdots(n_{i}-q+1)}{\bar{\epsilon_{i}}^{q}}\rangle}{\langle\frac{1}{M^{2}}\sum_{i=1}^{M^{2}}\frac{n_{i}}{\bar{\epsilon_{i}}}\rangle^{q}}.
 \label{Eq:FMcorrection}
 \end{eqnarray}

\noindent Where $n_{i}$ is the number of particles located in the $i$-th cell. $\bar{\epsilon_{i}}$ represents the event average of the mean efficiency of all the particles in the $i$-th cell, which is calculated by $\langle\frac{\sum_{j=1}^{n_i}\epsilon_i^j}{n_i}\rangle$. The efficiency correction of Eq.~\eqref{Eq:FMcorrection} is known as the cell-by-cell method~\cite{RefEfficiency}.

\begin{figure}[!htp]
\centering
   \includegraphics[scale=0.60]{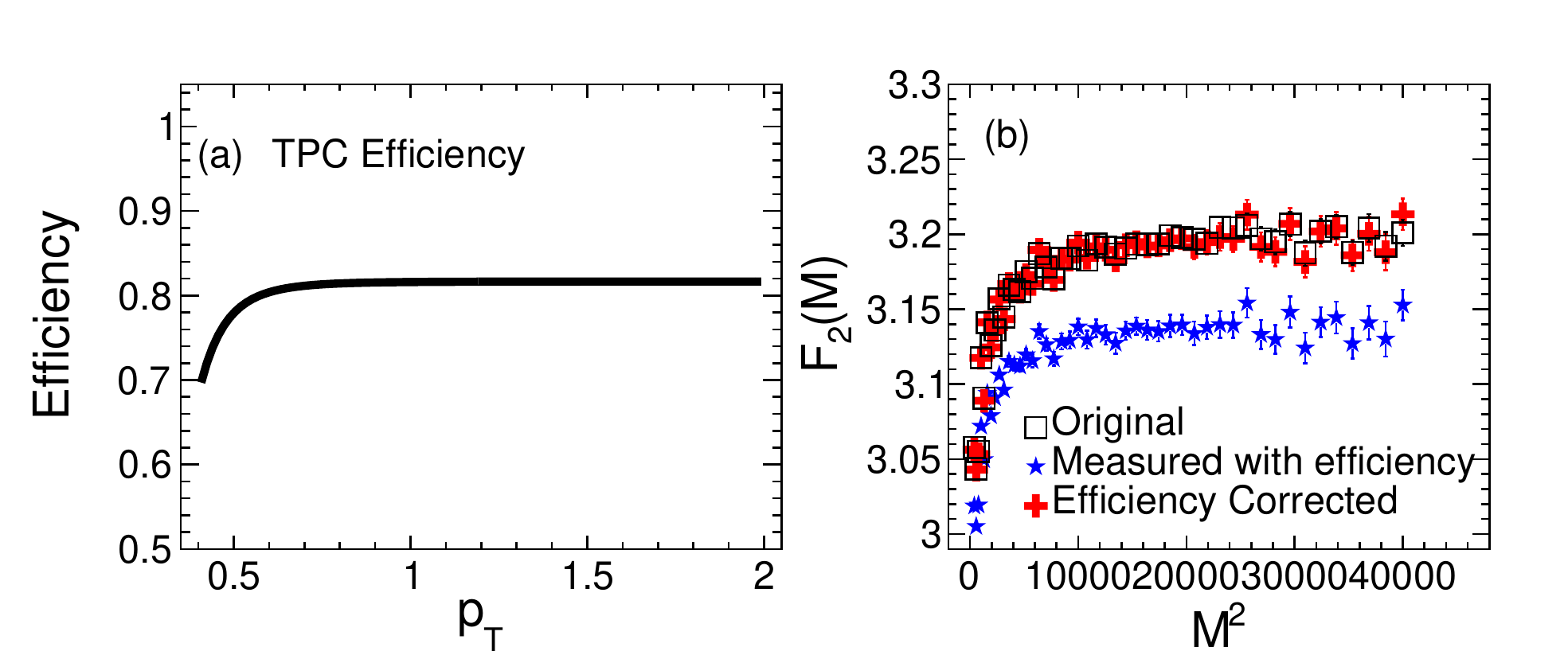}
   \caption{(a) The STAR experimental detection efficiency of protons as a function of $p_{T}$ in the TPC detector in Au+Au collisions at $\sqrt{s_\mathrm{NN}}$ = 19.6 GeV. (b) The original true $F_{2}(M)$ (open squares) as a function of number of bins in the transverse momentum region, together with the measured $F_{2}(M)$ (solid stars) after discarding particles and efficiency corrected SFM (solid crosses) by using the proposed cell-by-cell method.}
 \label{Fig:F2TPCEfficiency}
 \end{figure} 

\begin{figure}[hbtp]
\centering
\includegraphics[scale=0.60]{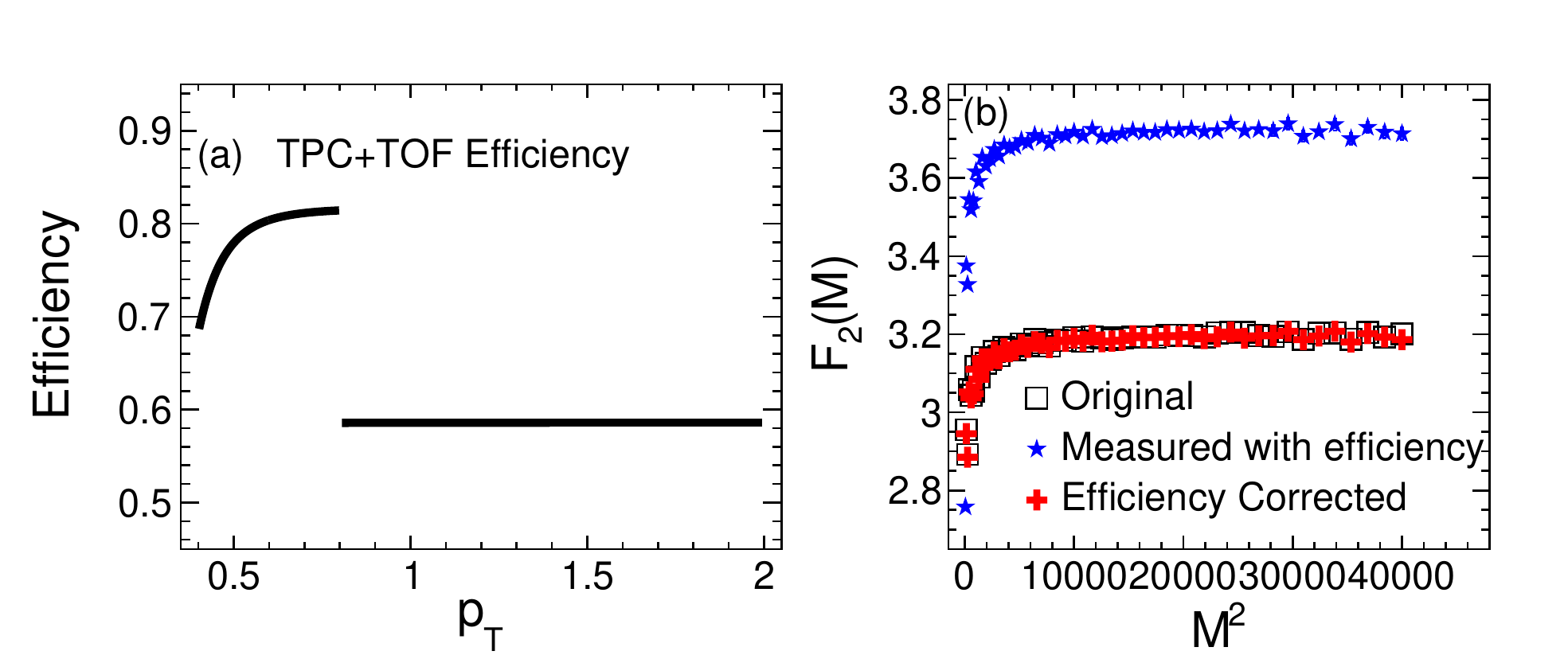}
  \caption{(a) The STAR experimental tracking efficiency as a function of $p_{T}$ in the TPC+TOF detectors in Au+Au collisions at $\sqrt{s_\mathrm{NN}}$ = 19.6 GeV. (b) The original true $F_{2}(M)$ (open squares) as a function of $M^2$, together with the measured $F_{2}(M)$ (solid stars) after discarding particles and the efficiency corrected $F_{2}(M)$ (solid crosses).}
  \label{Fig:F2TPTOFEfficiency}
\end{figure}   

In order to test the accuracy of the cell-by-cell method, we apply the tracking efficiencies in the real experiment to the event samples generated by the transport UrQMD model~\cite{RefEfficiency}. The experimental detection efficiency of tracks can be gained from Monte-Carlo embedding technique which considering effects including detector efficiency, acceptance, decays and interaction losses in the collisions~\cite{STARData,STARPRCMoment}. In the STAR experiment at RHIC, the efficiency of the Time Projection Chamber (TPC) or the  Time-of-Flight (TOF) detector is a function of transverse momentum~\cite{STARData,STARPRCMoment}. Figure~\ref{Fig:F2TPCEfficiency} (a) and~\ref{Fig:F2TPTOFEfficiency} (a) show the detection efficiencies of protons as a function of $p_{T}$ in the TPC detector and in the TPC+TOF detectors, respectively.

The tracking efficiency is then employed into the UrQMD event samples by keeping or discarding a particle based on the probability obtained from the $p_{T}$-dependent efficiencies. Figure~\ref{Fig:F2TPCEfficiency} (b) shows the original true $F_{2}(M)$ (open squares), measured $F_{2}(M)$ (solid stars) after discarding particles and efficiency corrected SFM (solid cross) calculated by the cell-by-cell method, respectively. We observe that the values of the measured $F_{2}(M)$ are smaller than those of the original true ones, especially in the region of large $M^{2}$. However, the efficiency corrected $F_{2}(M)$ agree well with the original true ones. Fig.~\ref{Fig:F2TPTOFEfficiency} (b) shows the calculations for the case of TPC+TOF detection efficiency. The corrected SFMs are confirmed to be consistent with the original true ones. 

Therefore, the proposed cell-by-cell method provides an accurate and effective way of efficiency corrections in the measurement of SFMs in heavy-ion collisions. This method has been adopted recently in the intermittency analysis in the STAR experiment~\cite{STARIntermittency}.  

\section{Search for Intermittency in Experiments}
The exploration of experimental evidence of creation of the QGP and location of the CP is one of the major objectives in current heavy-ion collisions. The goal is pursued by measurements of correlations and fluctuations including intermittency, which are supposed to be promising signatures of the QCD phase transition.

\subsection{Results from NA49 and NA61 Collaborations at SPS}
In the experiments at SPS, the search for the QCD critical point is attempted by NA49 experiment through changing system sizes of colliding nuclei (p+p, C+C, Si+Si, Pb+Pb) at 158$A$ GeV/$c$ and NA61/SHINE experiment by varying energies in p+p,p+Pb, Be+Be, Ar+Sc and Xe+La collisions.

\begin{figure}[htp]
     \centering
     \includegraphics[scale=0.4]{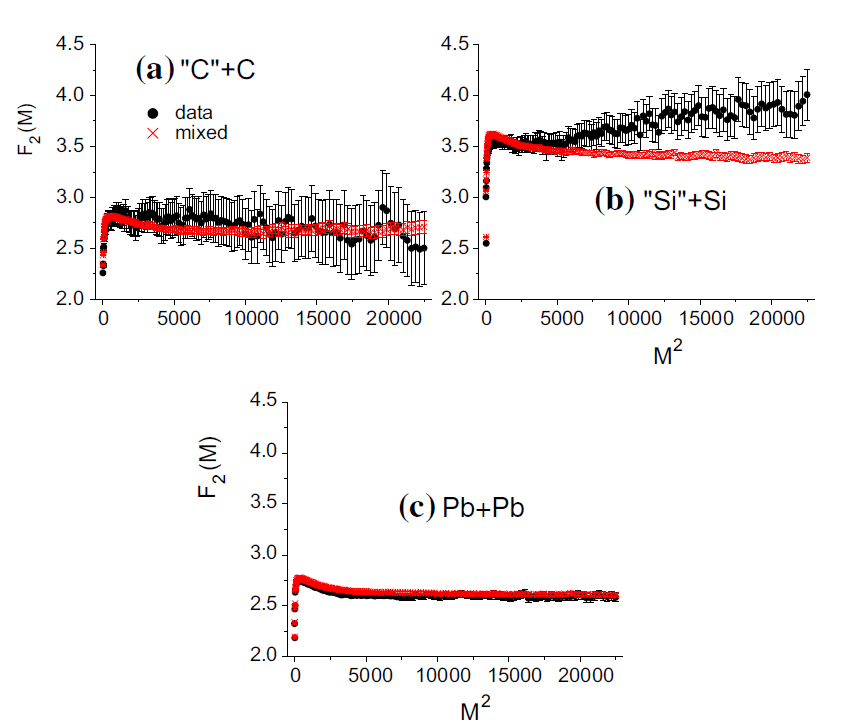}
     \caption{The measured $F_{2}(M)$ as a function of $M^{2}$ of proton density in the most central collisions at $\sqrt{s_\mathrm{NN}}$ = 17.3 GeV for (a) C+C, (b) Si+Si, and (c) Pb+Pb collisions. Figure taken from Ref~\cite{NA49EPJC}.}
     \label{FigNA49SiCPb}
\end{figure}

Figure~\ref{FigNA49SiCPb} presents $F_{2}(M)$ as a function of $M^{2}$ for data and mixed events at $\sqrt{s_\mathrm{NN}}$ = 17.3 GeV in C+C, Si+Si and Pb+Pb collisions measured in NA49 experiment. It is observed that $F_{2}(M)$ in the Si+Si collisions are larger than those calculated from mixed events when $M^{2}$ is large. The correlator $\Delta F_{2}(M)$ of Si+Si collision calculated by Eq.~\ref{Eq:DeltaFq} is found to obey a good power-law dependence on $M^{2}$ with $\phi_{2}=0.96\pm0.16$ which approaches theoretical prediction~\cite{AntoniouPRL}, indicating a typical property of intermittency in this collision. However, the $F_{2}(M)$ are almost overlap with those of mixed events in the C+C and Pb+Pb systems, suggesting that intermittency is not visible in these collisions. The reason could be that the critical self-similar fluctuations can not develop in the small size of C+C system. As for the Pb+Pb collision, the signal might be diluted during the longer evolution of the hadronic phase~\cite{NA49EPJC}.

\begin{figure}[htp]
     \centering
     \includegraphics[scale=0.25]{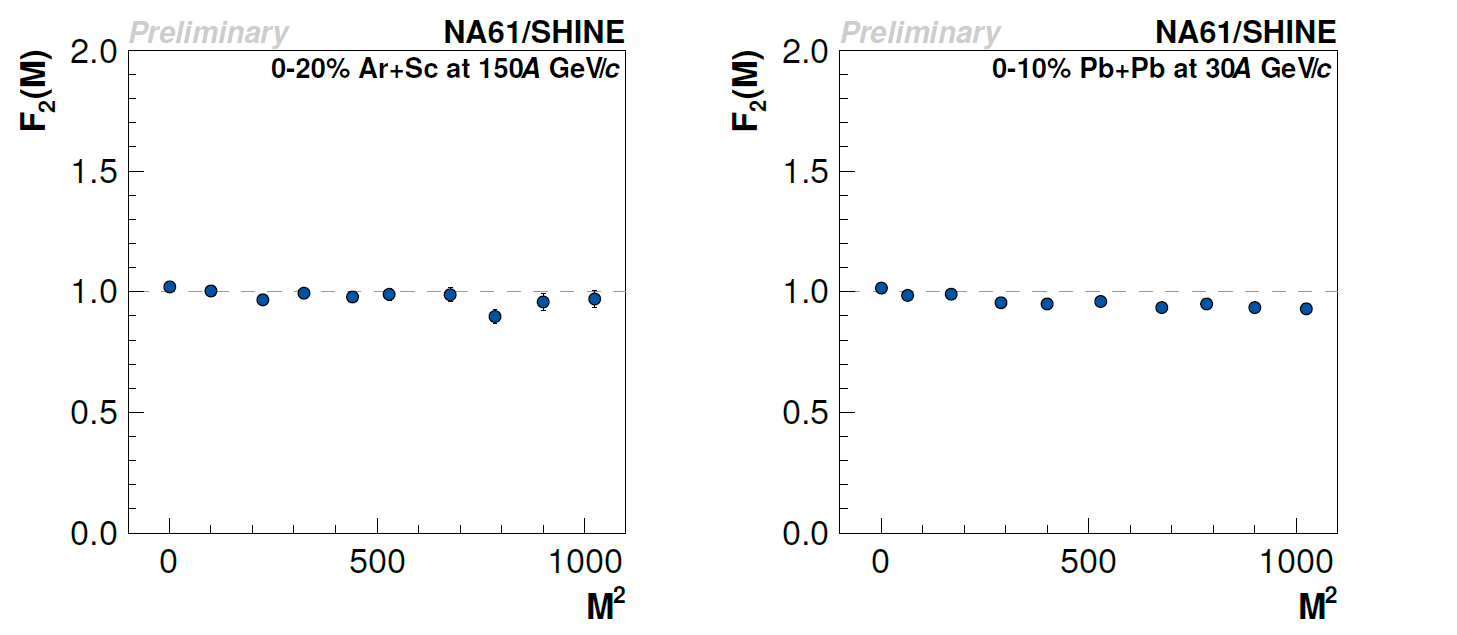}
     \caption{The measured $F_{2}(M)$ as a function of $M^{2}$ of proton density in 0-20\% central Ar+Sc collisions at 150$A$ GeV/$c$ (left pad) and in 0-10\% central Pb+Pb collisions at 30$A$ GeV/$c$ (right pad). Figure taken from Ref~\cite{NA61CFM}.}
     \label{FigNA61ArPb}
\end{figure}

Figure~\ref{FigNA61ArPb} illustrates the preliminary results from NA61 on the second-order SFMs of proton numbers in 0-20\% central Ar+Sc collisions at 150$A$ GeV/$c$ and in 0-10\% central Pb+Pb collisions at 30$A$ GeV/$c$~\cite{NA61CFM}. The cumulative variable method is used here to subtract background effects of the measurement. It is observed that $F_{2}(M)$ is nearly flat with increasing $M^{2}$, indicating the absence of an intermittency behavior in Ar+Sc or in Pb+Pb collisions. 
   
\subsection{Results from STAR Experiment at RHIC}
To explore the QCD phase diagram, the STAR Collaboration has measured a few experimental observables which may signature the QCD phase transition and CP at RHIC BES-I energies. In particularly, some appealing non-monotonic behaviors of the measurements have been observed in Au+Au collisions at energies around $20<\sqrt{s_{NN}}<30$ GeV~\cite{STARPRLMoment,HBTSTAR,HBTPRL,STARLightRatio,STARflowV1}.

\begin{figure*}[!htp]
     \centering
     \includegraphics[scale=0.70]{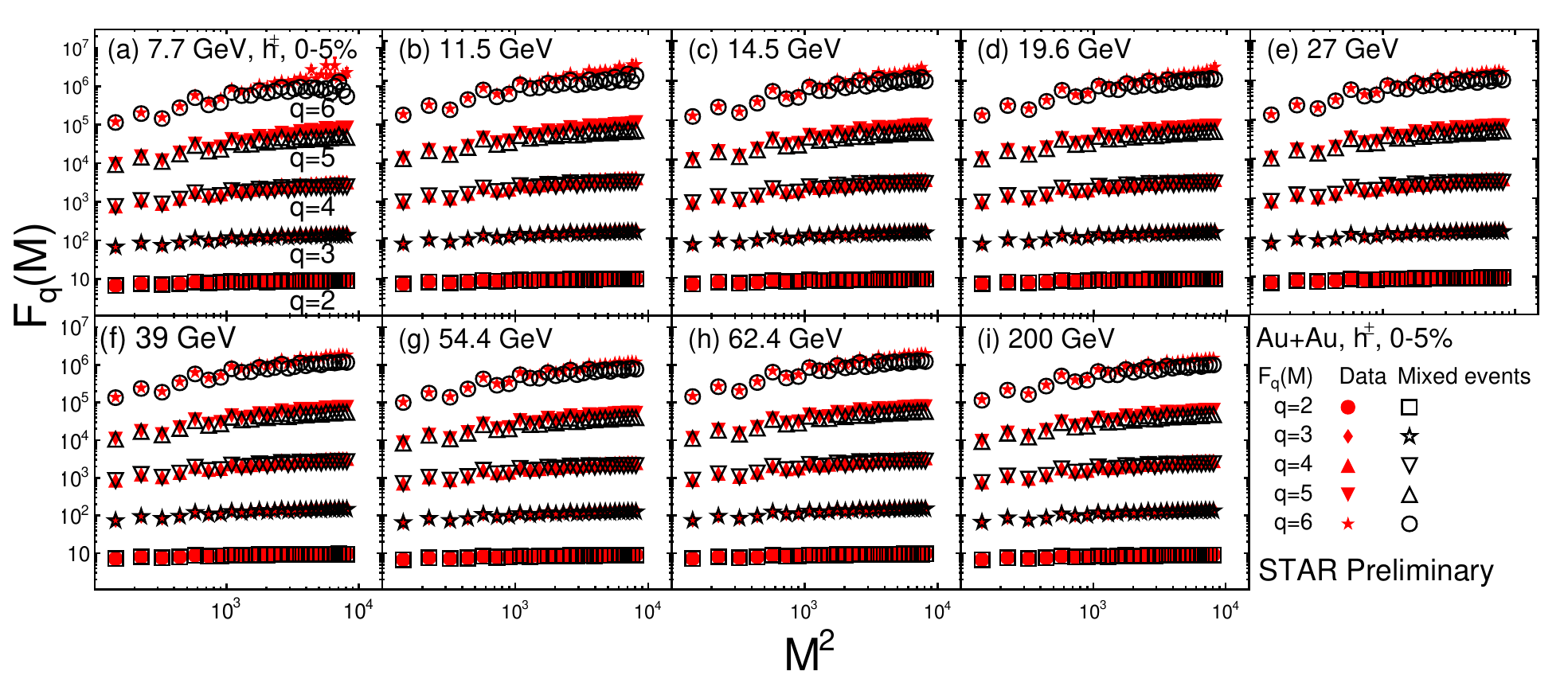}
     \caption{$F_{q}(M)$ (q=2-6) as function of $M^{2}$ for charged particles in the 0-5\% most central Au+Au collisions at $\sqrt{s_\mathrm{NN}}$ = 7.7-200 GeV in double-logarithmic scale. Red (black) marks represent $F_{q}(M)$ of data (mixed event) as a function of $M^{2}$. Figure taken from Ref~\cite{STARIntermittency}.}
     \label{FigSFMSTAR}
\end{figure*}

Recently, the STAR Collaboration reports the preliminary results on the first measurements on intermittency of charged particles in Au+Au collisions at $\sqrt{s_\mathrm{NN}}$= 7.7-200 GeV~\cite{STARIntermittency}. The SFMs of $p$, $\bar{p}$, $K^{\pm}$ and $\pi^{\pm}$ have been measured within $|\eta|<0.5$ in a 2D transverse momentum space. The SFMs are corrected for the finite tracking reconstruction efficiencies by the cell-by-cell method. Figure~\ref{FigSFMSTAR} shows the $F_{q}(M)$ of both the data (solid symbols) and the mixed events (open symbols) for identified charged particles in the $0-5\%$ most central Au+Au collision at BES-I energies. The $F_{q}(M)$ can be calculated up to the sixth-order based on the event statistics. It is observed that $F_{q}(M)^{data}$ is larger than $F_{q}(M)^{mix}$ at the region of large $M^{2}$ values at various energies and thus a deviation of $\Delta F_{q}(M)$ from zero is found in these collisions.

\begin{figure*}[!htp]
     \centering
     \includegraphics[scale=0.40]{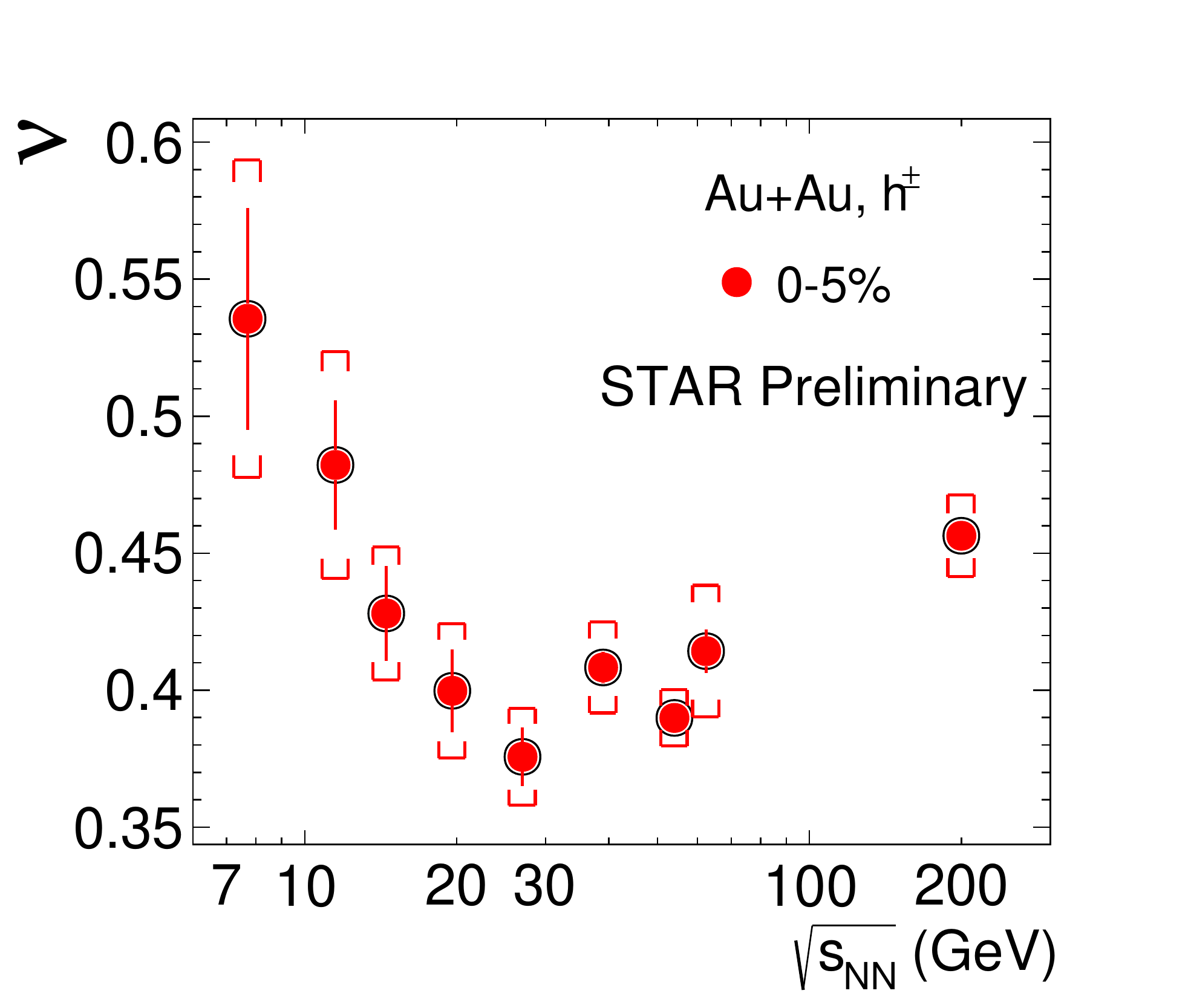}
     \caption{Energy dependence of the scaling exponent $\nu$ for charged particles in Au+Au collisions at $\sqrt{s_\mathrm{NN}}$ = 7.7-200 GeV. Figure taken from Ref~\cite{STARIntermittency}.}
     \label{Fig:nuenergy}
\end{figure*}

After subtracting the background contributions, the critical related exponent $\nu$, which extracted from the $\Delta F_{q}(M)/\Delta F_{2}(M)$ scaling according to Eq.~\eqref{Eq:FqF2scaing} and \eqref{Eq:nuscaing}, exhibits a non-monotonic behavior on energy with a dip located between 20 GeV and 30 GeV in the most central Au+Au collisions, as shown in Fig.~\ref{Fig:nuenergy}. The non-monotonic energy dependence of $\nu$ needs to be understood with more theoretical inputs, especially a reliable theoretical calculation in a 2D transverse momentum space with the same acceptance cuts used in the experimental analysis.
   
\section{Intermittency Analysis in Models}

\subsection{Scaled Factorial Moments of Protons in the Transport UrQMD Model} 
In this section, we illustrate the SFMs of proton density calculated in different energies and centralities by using the UrQMD model. The UrQMD (Ultra relativistic Quantum Molecular Dynamics) is a transport model~\cite{UrQMD1, UrQMD2} which is widely used to simulate relativistic heavy-ion collisions with a wide energy coverage ranging from SIS energies to top of the RHIC energy. Since the model does not include any correlations or fluctuations related to the QCD phase transition or CP, it is feasible to be used to study background effects or other trivial fluctuations in the search of the QCD CP in heavy-ion collisions.
 
The cascade mode of the UrQMD model with the version 2.3 is applied to produce events in Au+Au collision at $\sqrt{s_\mathrm{NN}}$ = 7.7, 11.5, 19.6, 27, 39, 62.4 and 200 GeV. In the model calculation, we apply the same kinematic cuts and analysis methods as those used in the STAR experiment~\cite{STARPRCMoment}. Protons are selected at mid-rapidity($|y|<0.5$) in transverse momentum space $0.4<p_{T}<2.0$. Centrality is defined by charged pions and kaons in pseudo-rapidity region $|\eta|<1.0$.
  
\begin{figure*}
     \centering
     \includegraphics[scale=0.70]{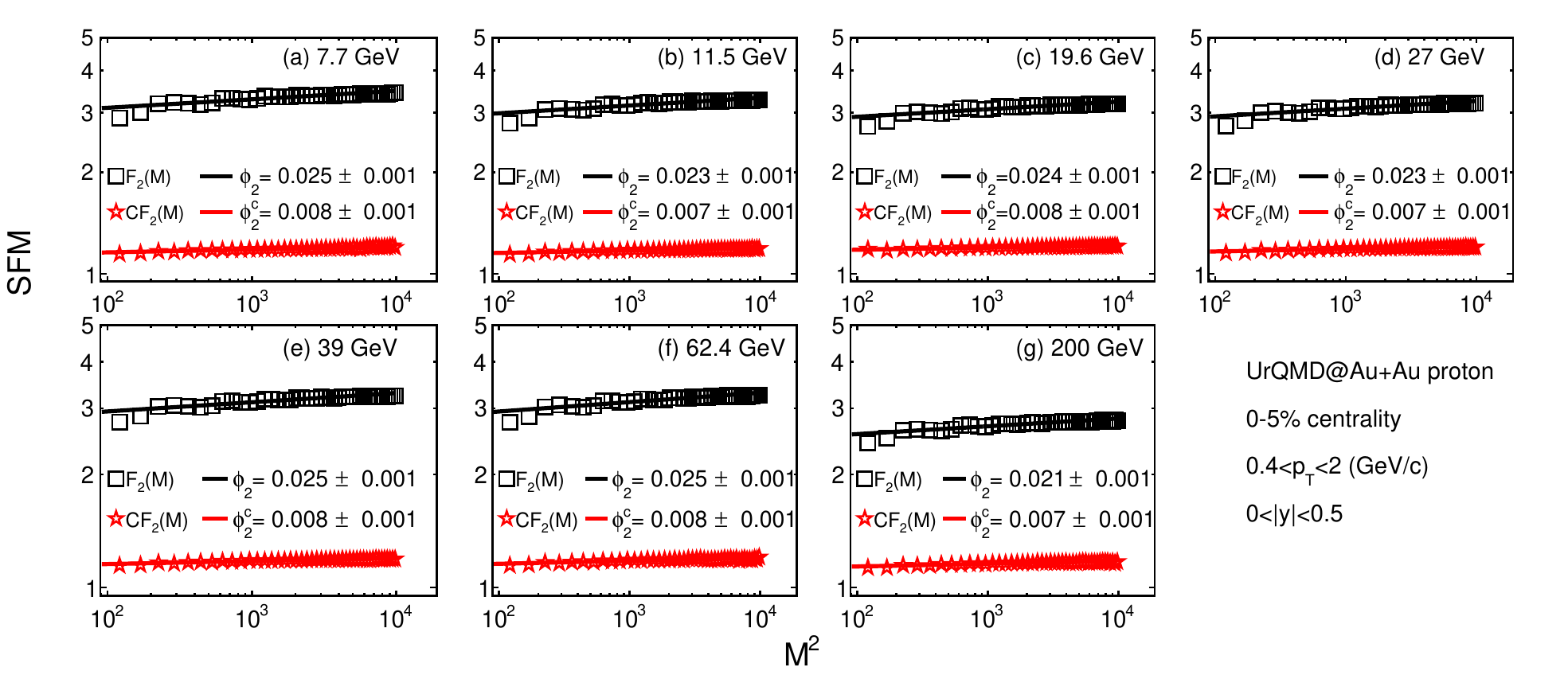}
      \caption{The second-order SFM (black squares) as a function of $M^{2}$ in the 0-5\% most central Au+Au collisions at $\sqrt{s_\mathrm{NN}}$ = 7.7-200 GeV. The black solid lines are the power-law fitting according to Eq.~\eqref{Eq:PowerLaw}. The corresponding red ones are those calculated by the cumulative variable method.}
     \label{Fig:F2EUrQMD}
 \end{figure*}
 
 \begin{figure*}
  \centering 
  \includegraphics[scale=0.70]{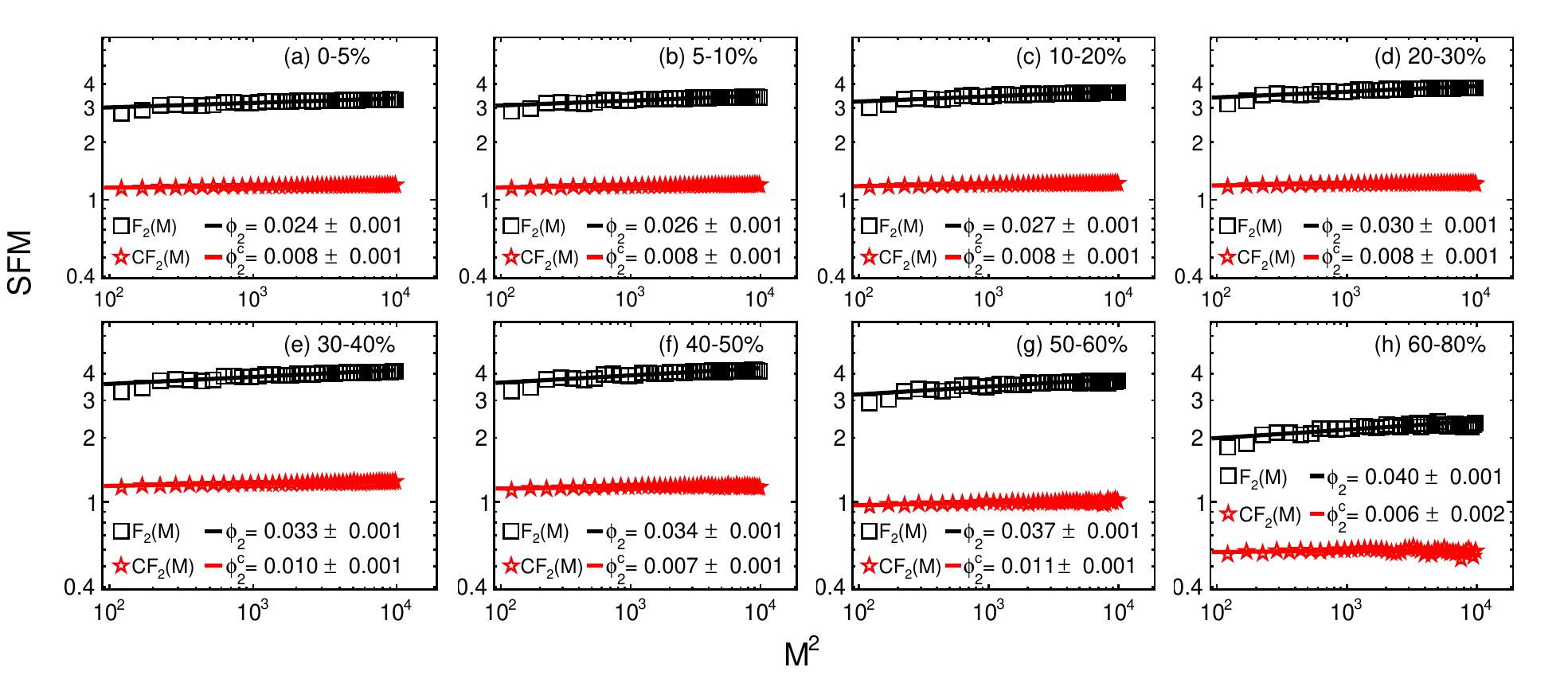}
  \caption{The second-order SFM (black squares) as a function of $M^{2}$ measured in various collision centralities in Au+Au collisions at $\sqrt{s_\mathrm{NN}}$ = 19.6 GeV. The corresponding red ones are those calculated by the cumulative variable method.}
  \label{Fig:F2CentUqMD}
\end{figure*}
 
Figure~\ref{Fig:F2EUrQMD} shows the directly measured second-order SFMs (black squares) and the ones with background subtraction by the cumulative variable method (red stars), as a function of the number of partitioned cells for protons in the 0-5\% central Au+Au collisions at $\sqrt{s_\mathrm{NN}}$ = 7.7-200 GeV. It is found that directly measured $F_{2}(M)$ increases slowly with increasing $M^{2}$. By using the cumulative method to subtract the background contribution, the $CF_{2}(M)$ shows a nearly flat trend with $M^{2}$. The value of $\phi_{2}^{c}$, which is fitted from SFMs by the cumulative variable method, is found to be near zero at the measured energies.

In Fig.~\ref{Fig:F2CentUqMD}, we show the $F_{2}(M)$ and $CF_{2}(M)$ at $\sqrt{s_\mathrm{NN}}$ = 19.6 GeV in various collision centralities. The values of $\phi_{2}$ calculated from directly measured SFMs slightly rise from the most central to the most peripheral collisions. However, $\phi_{2}^{c}$ calculated from $CF_{2}(M)$ show a flat centrality dependence with all the values near zero. The results verify that the background of the non-critical contributions can be efficiently wiped off by using the proposed cumulative variable
method in the UrQMD model.   

\subsection{Intermittency in the CMC Model belonging to the 3D Ising Universality Class}
Based on an effective action of the three-dimensional Ising universality class, there exists a power-law singularity of density-density correlator for small momentum transfer $\vec{k}$~\cite{AntoniouPRL}:
\begin{equation}
\lim\limits_{|\vec{k}|\rightarrow 0}\langle\rho_{\vec{k}}\rho_{\vec{k}}^{*}\rangle\sim |\vec{k}|^{-d_{F}}.
 \label{Eq:CorrMom}
\end{equation}

\noindent Here, $\rho_{\vec{k}}$ represents the baryon-number density in momentum space by Fourier transforms from the one in coordinate space. $\langle\rho_{\vec{k}}\rho_{\vec{k}}^{*}\rangle$ is the density-density correlator of particle pairs in momentum space. Eq.~\eqref{Eq:CorrMom} shows a fractal geometry with a power-law pattern in momentum space for critical systems belonging to the 3D Ising universality class.

\begin{figure}[!htb]
\centering
\includegraphics[scale=0.4]{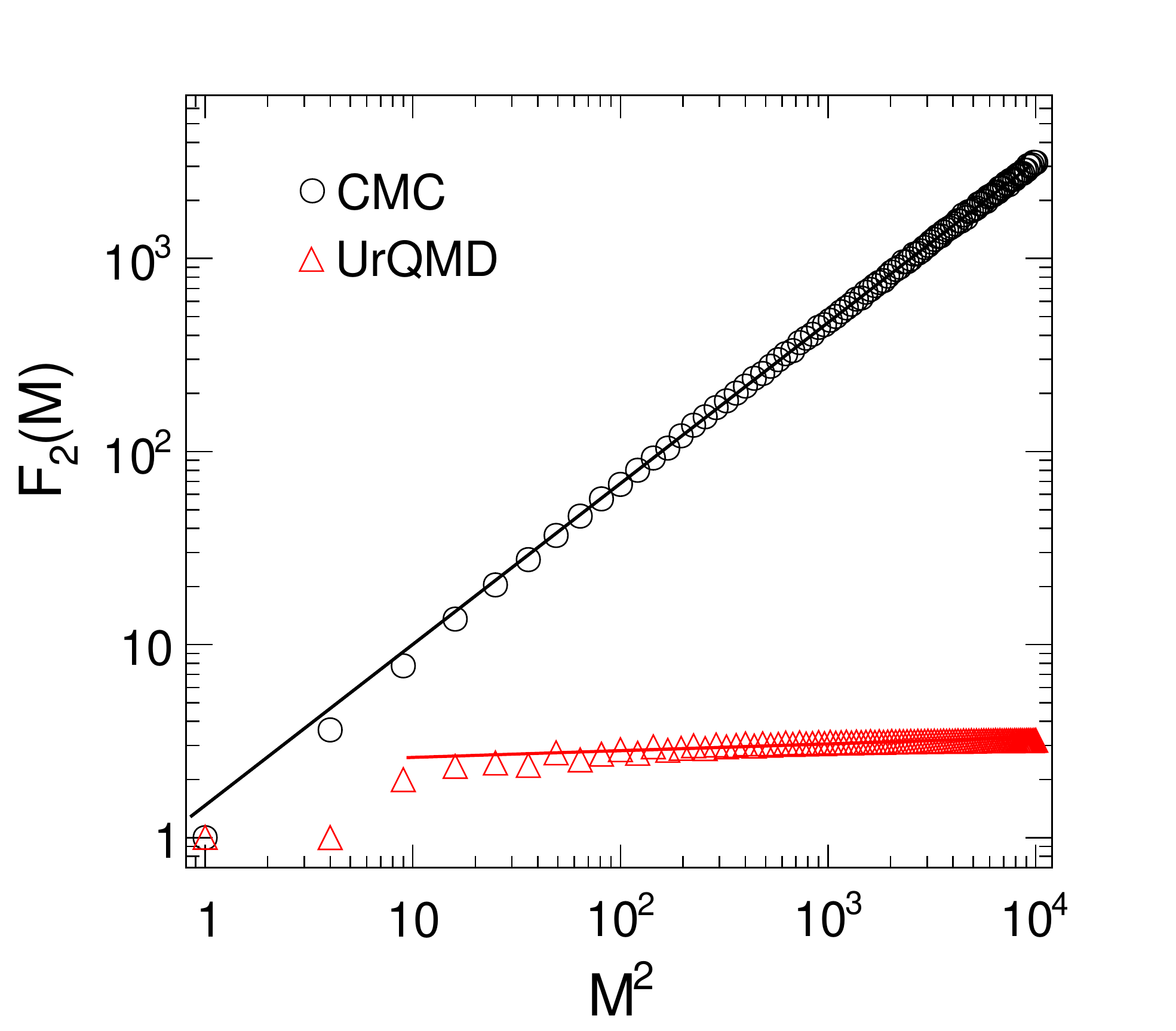}
\caption{$F_{2}(M)$ as a function of $M^{2}$ from the CMC events (black circles) and from the UrQMD model (red triangles) in Au+Au collisions at $\sqrt{s_\mathrm{NN}}$ = 19.6 GeV in a double-logarithmic scale. Figure taken from Ref~\cite{CMCPLB}.}
\label{FigCMCandUrQMD}
\end{figure}

In the study on intermittency in heavy-ion collisions, the CMC model~\cite{AntoniouPRL} is widely used to simulate critical event samples incorporating self-similar fluctuations. The Monte-Carlo simulations involving critical density fluctuations related to Eq.~\eqref{Eq:CorrMom} can be simulated by the algorithm of Levy random walk, which requiring the probability density $\rho(p)$ between two adjacent walks follows:
\begin{equation}
\rho(p)=\frac{\nu p_{\rm min}^{\nu}}{1-(p_{\rm min}/p_{\rm max})^{\nu}}p^{-1-\nu}.
 \label{Eq:probLevy}
\end{equation}

\noindent Here $\nu$ is the Levy exponent directly connected to the intermittency index, $p$ represents the momentum distance of two adjacent particles, $p_{\rm min}$ and $p_{\rm max}$ are the minimum and maximum values of $p$. The parameters of Levy function are set to be $\nu = 1/6$ and $p_{\rm min}/p_{\rm max} = 10^{-7}$ for critical events leading to the second-order intermittency $\phi_{2}=\frac{5}{6}$. The precise rules of the Levy algorithm and more details about the CMC model could be found in Refs~\cite{PRELevy,AntoniouPRL}.

In Fig.~\ref{FigCMCandUrQMD}, the black circles show $F_{2}(M)$ as a function of $M^{2}$ and the solid black line is the fitting according to Eq.~\eqref{Eq:PowerLaw}. The results are calculated from a generated sample of 600 critical events. The Levy algorithm is parameterized to produce $p_{x}$ and $p_{y}$ distributions of particles in transverse momentum space and the multiplicity is set to obey a Poisson distribution with the mean value $\langle N\rangle = 20$. It is found that $F_{2}(M)$ from CMC events does obey a good power-law dependence on $M^{2}$ with the fitting slope $\phi_{2}=0.834\pm0.001$. It infers that the CMC model could excellently reproduce the intermittency behavior related to the self-similar correlations of Eq.~\eqref{Eq:CorrMom}. The open red triangles are the results calculated from the UrQMD event sample by the same mean multiplicity as used in the CMC model. It gives a flat trend of the dependence on $M^{2}$. The reason is that this model does not incorporate any self-similar fluctuations in the mechanism of particle production process. 

The definition of the relative density fluctuation of baryons, $\Delta n$, is ~\cite{KJSunPLB2017,KJSunPLB2018}
\begin{equation}
\Delta n=\frac{\langle(\delta n)^2\rangle}{\langle n\rangle^2}=\frac{\langle n^2\rangle-\langle n\rangle^2}{\langle n\rangle^2}.
 \label{Eq:relativeDF}
\end{equation}

\noindent The angle bracket represents the average over the whole event sample. 

\begin{figure}[!htb]
\centering
\includegraphics[scale=0.4]{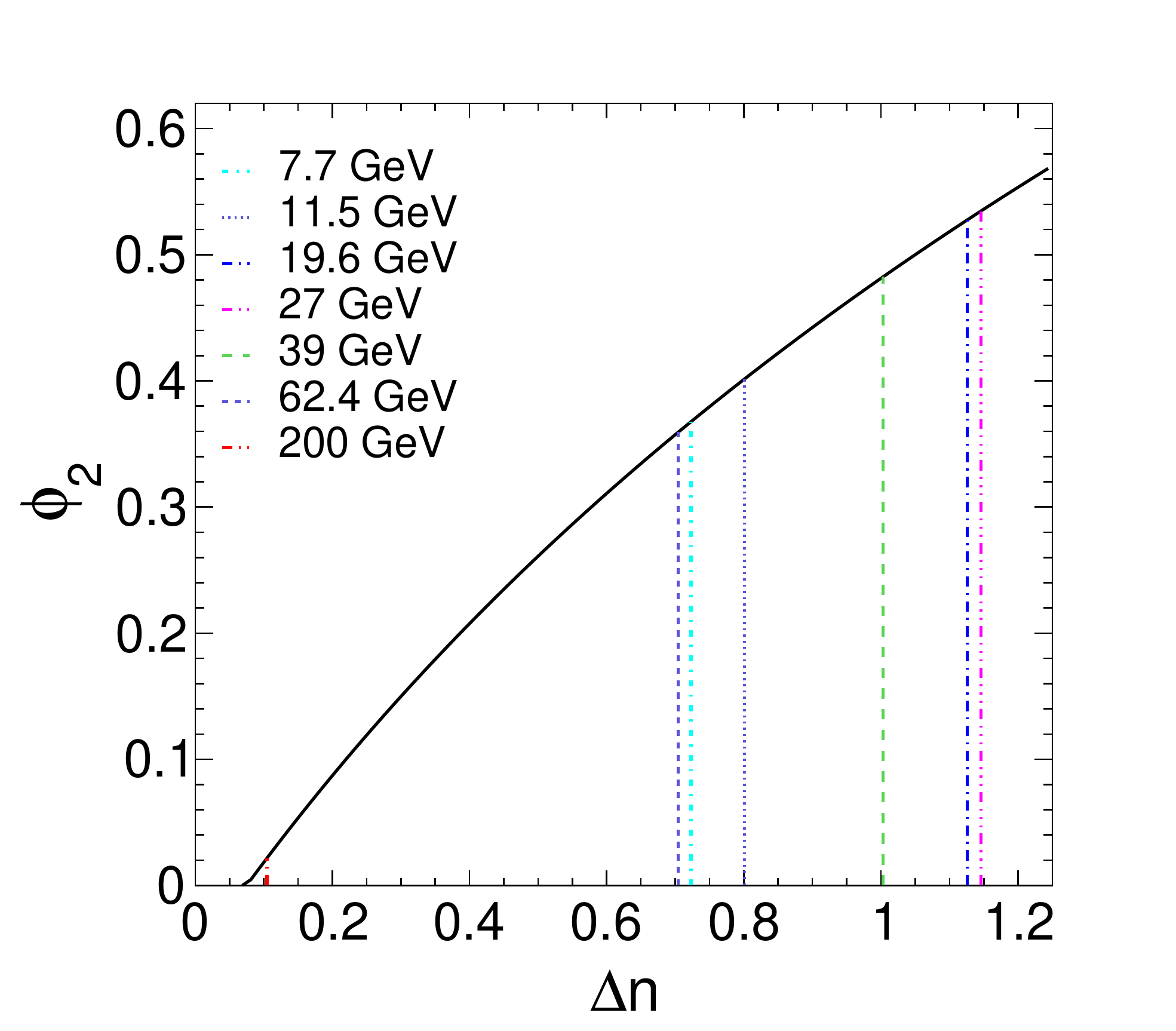}
\caption{The second-order intermittency index as a function of relative density fluctuation in the CMC model. The dash lines represent the experimental measured $\Delta n$ at $\sqrt{s_\mathrm{NN}}$ = 7.7-200 GeV from the STAR experiment~\cite{STARLightRatio}. Figure taken from Ref~\cite{CMCPLB}.}
\label{FigCMCPphiDeltan}
\end{figure}

Measurements of SFMs to get intermittency index and calculations of $\Delta n$ can be performed simultaneously in the same CMC event sample~\cite{CMCPLB}. The solid black line in Fig.~\ref{FigCMCPphiDeltan} displays the second-order intermittency index as a function of the relative density fluctuation. We observe a monotonically increase of $\phi_2$ on $\Delta n$. It reveals a fact that great intermittency can be achieved if large baryon density fluctuations are developed for the system near the QCD critical point. It supplies another experimentally measurable variable to obtain density fluctuations besides measurement of light nuclei productions in the coalescence model~\cite{KJSunPLB2017,KJSunPLB2018}. The colored dash lines in the same figure show the STAR measured $\Delta n$ in the 0-10\% central Au+Au collisions at $\sqrt{s_\mathrm{NN}}$ = 7.7, 11.5, 14.5, 19.6, 27, 39, 62.4 and 200 GeV, respectively~\cite{STARLightRatio}.

\begin{figure}[!htb]
\centering
\includegraphics[scale=0.40]{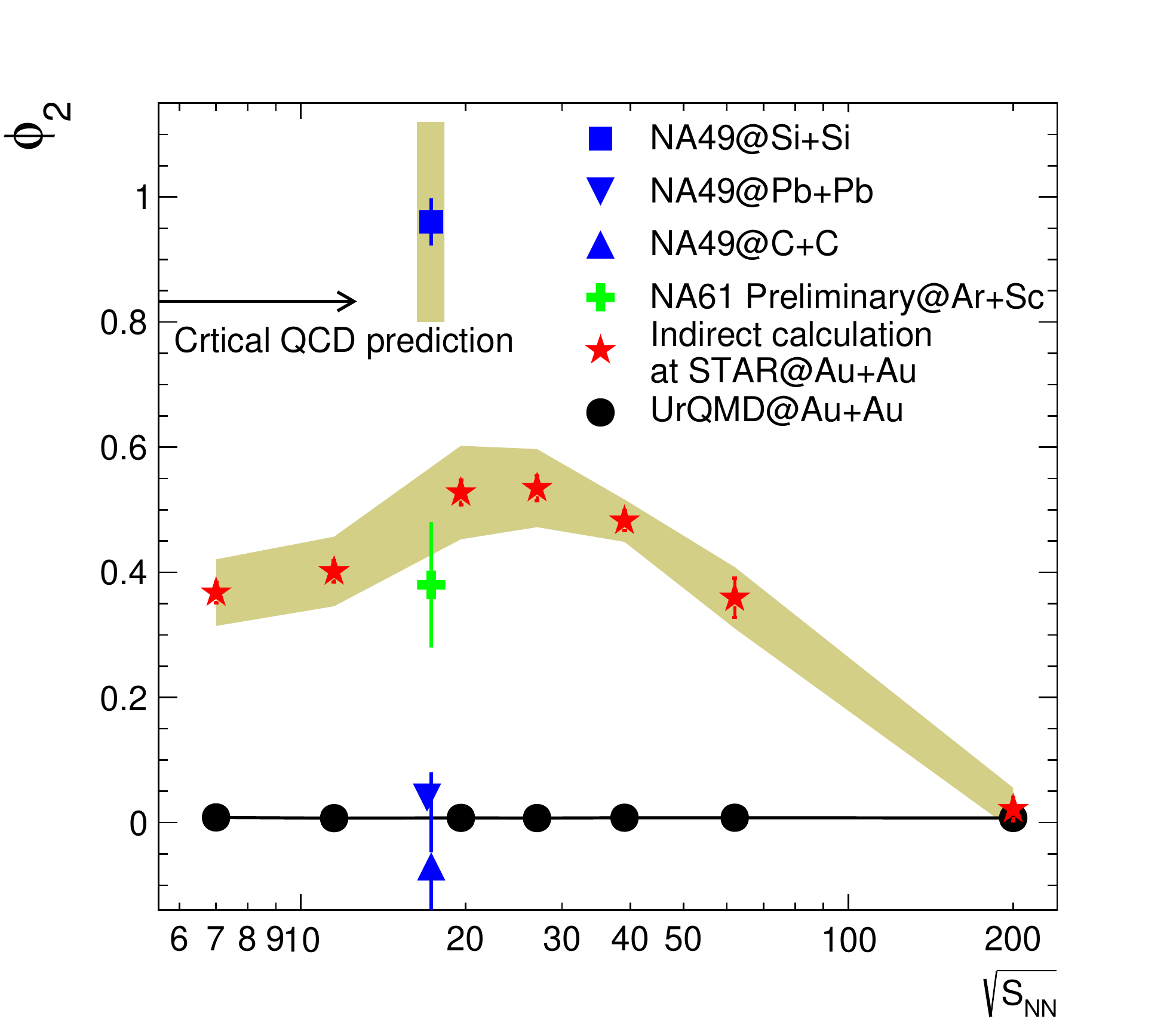}
\caption{The collision energy dependence of the intermittency index $\phi_{2}$ in various collision systems.  The solid blue and green symbols represent the values from the NA49~\cite{NA49PRC,NA49EPJC} and NA61 experiments~\cite{NGNPA2020,NA61universe}. The red stars are the $\phi_{2}$ indirectly extracted from the measured light nuclei productions in the STAR experiment~\cite{STARLightRatio}. The black circles show the intermittency index of protons from UrQMD model. The black arrow is the theoretic expectation for a critical model~\cite{AntoniouPRL}. }
\label{FigAllphiq}
\end{figure}

Once the connection between $\phi_{2}$ and $\Delta n$ is gained, the intermittency index can be obtained by measuring the density fluctuations
within the same event sample, or vice versa. In this way, we can calculate $\phi_{2}$ indirectly by mapping the experimental measured $\Delta n$ into the relation. In Fig.~\ref{FigAllphiq}, the red stars show the energy dependence of indirectly calculated $\phi_{2}$ in the 0-10\% central Au+Au collisions. We observe a non-monotonic dependence of $\phi_{2}$ on collision energy with a peak structure at around 20-30 GeV. It indicates that the strength of intermittency is strongest in this energy region. For comparison, data from the NA49~\cite{NA49PRC,NA49EPJC} (blue symbols) and the NA61/SHINE~\cite{NGNPA2020,NA61universe} (green crosses) experiments are also plotted in the same figure. In Si+Si collisions, the $\phi_{2}$ is found to be close to the value of theoretical expectation which is illustrated as the black arrow. The black circles represent $\phi_{2}$ from the UrQMD calculations, which give a flat energy dependence with all the values around zeros. It is because that no critical self-similar mechanism is implemented in this model.  

\subsection{Intermittency Analysis in the UrQMD Model with Hadronic Potentials}
As has been shown in Section 6.1, there is no intermittency behavior in the original UrQMD model after subtracting background contributions. Recently, attempts have been made to  give obvious correlations of proton pairs and intermittency by introducing hadronic potentials into the UrQMD model with the mean-field mode~\cite{LiPLBUrQMD}. 

By treating potentials of both the formed and performed hadrons from string fragmentation by the same means, the density-dependent potentials can be written as ~\cite{LiPLBUrQMD,HadronLiPLB}:
\begin{equation}
 U=\alpha(\frac{\rho_{h}}{\rho_{0}})+\beta(\frac{\rho_{h}}{\rho_{0}})^{\gamma},
 \label{Eq:HadronPotential}
\end{equation}
\noindent where $\rho_{h}$ represents the hadronic density, and $\rho_{0}$ is the nuclear matter saturation density.

The momentum-dependent term of hadronic potential is defined as:
\begin{equation}
U_{md}=\sum_{k=1,2}\frac{t^{k}_{md}}{\rho_{0}}\int d\bm{p'}\frac{f({\bm{r},\bm{p'}})}{1+[(\bm{p}-\bm{p'})/a_{md}^{k}]^{2}}.
\label{EqMomentumTerm}
\end{equation}
\noindent Here $t_{md}$ and $a_{md}$ are parameters. More details about the implementation of hadronic potentials into the model could be found in Refs~\cite{LiPLBUrQMD,HadronLiPLB}.
 
\begin{figure}[htp]
     \centering
     \includegraphics[scale=1.2]{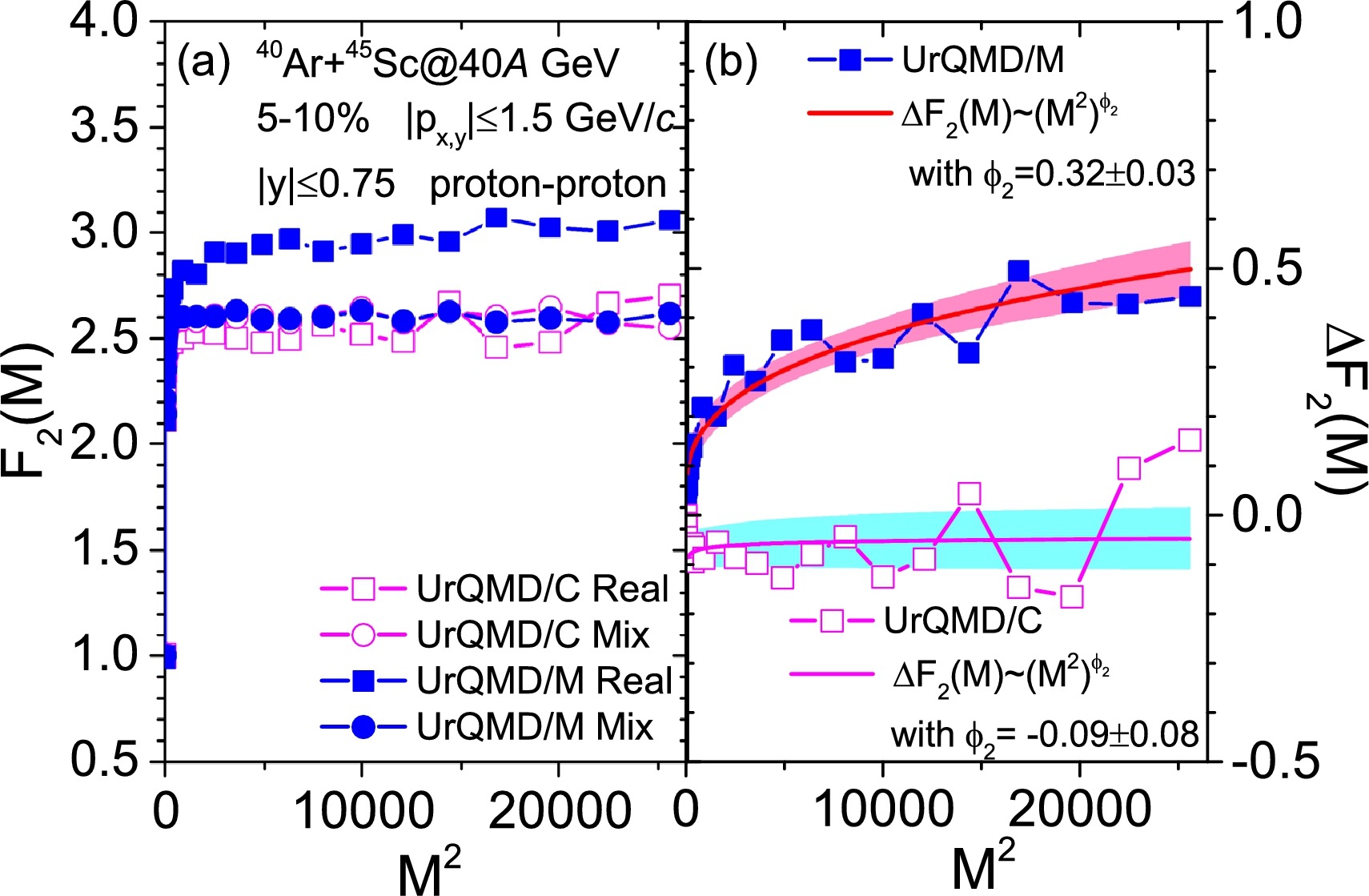}
     \caption{{(a)} The second-order SFM as a function of $M^{2}$ of proton density in the 5-10\% central Ar+Sc collisions at 40$A$ GeV/$c$. The solid squares (circles) are $F_{2}(M)$ of real events (mixed events) from the UrQMD model with hadronic potentials (UrQMD/M). The corresponding open squares and circles represent the measurements without hadronic potentials (UrQMD/C). {(b)} $\Delta F_{2} (M)$ as a function of $M^{2}$ of protons. Figure taken from Ref~\cite{LiPLBUrQMD}.}
     \label{FigUrQMDMandC}
\end{figure}

Figure~\ref{FigUrQMDMandC} illustrates the second-order $F_{2}(M)$ and the correlator $\Delta F_{2}(M)$ calculated in the UrQMD data samples both with and without hadronic potentials in Ar+Sc collisions at 40A GeV/$c$. As shown in the left figure of the UrQMD model with hadronic potential (UrQMD/M), $F_{2}(M)$ of original events (solid blue squares) are larger than those of mixed events (solid blue circles) at the large $M^{2}$ regions. In Fig.~\ref{FigUrQMDMandC} (b), the correlator $\Delta F_{2}(M)$ of hadronic potentials shows a power-law behavior with $M^{2}$. The extracted $\phi_{2}=0.32\pm 0.03$ is similar to the one measured in the NA61 experiment~\cite{NA61Davis,NA61universe}. As for the UrQMD cascade model (UrQMD/C), the values of $F_{2}(M)$ of real events are almost overlapped with those of mixed events and thus $\Delta F_{2}(M)$ is around 0 in this case. It could infer that the power-law character of $\Delta F_{2}(M)$ in the UrQMD/M model is introduced by the hadronic interaction, particular nuclear potentials which cause enhancements of proton pairs with small relative momenta~\cite{LiPLBUrQMD}.

\begin{figure}[htp]
     \centering
     \includegraphics[scale=1.2]{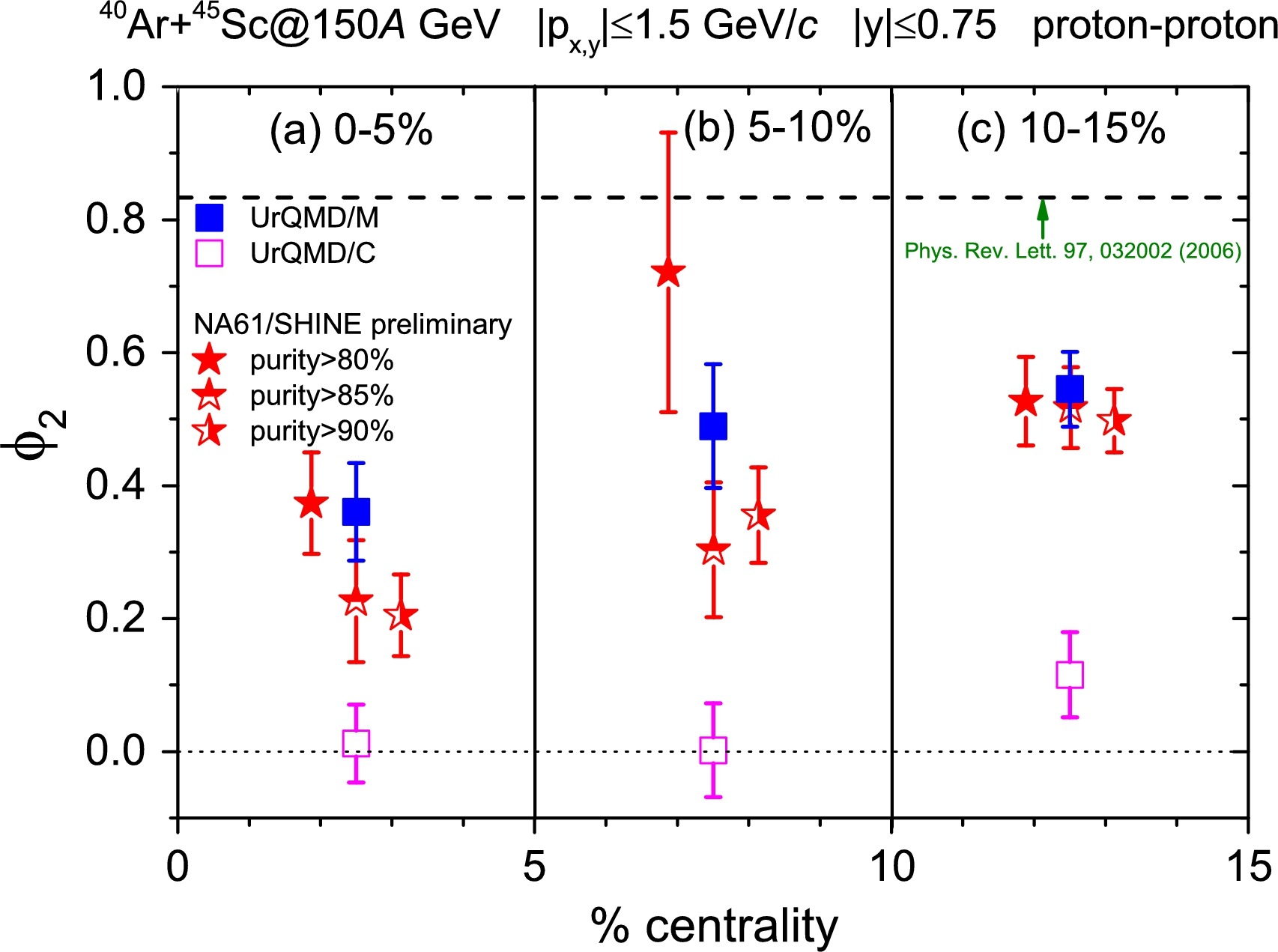}
     \caption{The extracted intermittency index $\phi_{2}$ at different centralities for Ar+Sc collisions in the UrQMD/M and UrQMD/C models. Figure taken from Ref~\cite{LiPLBUrQMD}.}
     \label{FigUrQMDMCentrality}
\end{figure}
    
Figure~\ref{FigUrQMDMCentrality} demonstrates centrality dependence of $\phi_{2}$ in 0-5\%, 5-10\%, and 10-15\% central Ar+Sc collisions, respectively. The calculations from the UrQMD model with (solid blue squares) and without (open pink squares) hadronic potential are plotted along with the NA61 preliminary experimental results (solid red stars). The second-order intermittency index $\phi_{2}$ calculated in the UrQMD/C model is found to be nearly zero. For the UrQMD/M model, the $\phi_{2}$ slightly increases from the most central to the mid-central collisions, in coincide with the experimental result for the cases of purity $>85\%$~\cite{NA61Davis,NA61universe}. This centrality dependence could be explained by a shortened hadronic freeze-out phase with decreasing size of the system~\cite{LiPLBUrQMD}.   

\subsection{Observation of Intermittency in the Hybrid UrQMD-hydro Model}
Apart from the introduce of hadronic potentials into the transport UrQMD model to obtain intermittency, another attempt is the hybrid UrQMD-hydro model which incorporating both transport and hydrodynamical descriptions of heavy-ion collisions~\cite{GopeEPJA,UrQMDhydro}. Hydrodynamic description has been suggested as an effective tool to express the hot and dense stage of collision reactions and model phase transitions~\cite{UrQMDhydro}. In this study, the intermediate hydrodynamic calculation is applied to the microscopic transport computation in initial condition and freeze-out process~\cite{UrQMDhydro,UrQMD2}.

\begin{figure}[htp]
     \centering
     \includegraphics[scale=0.35]{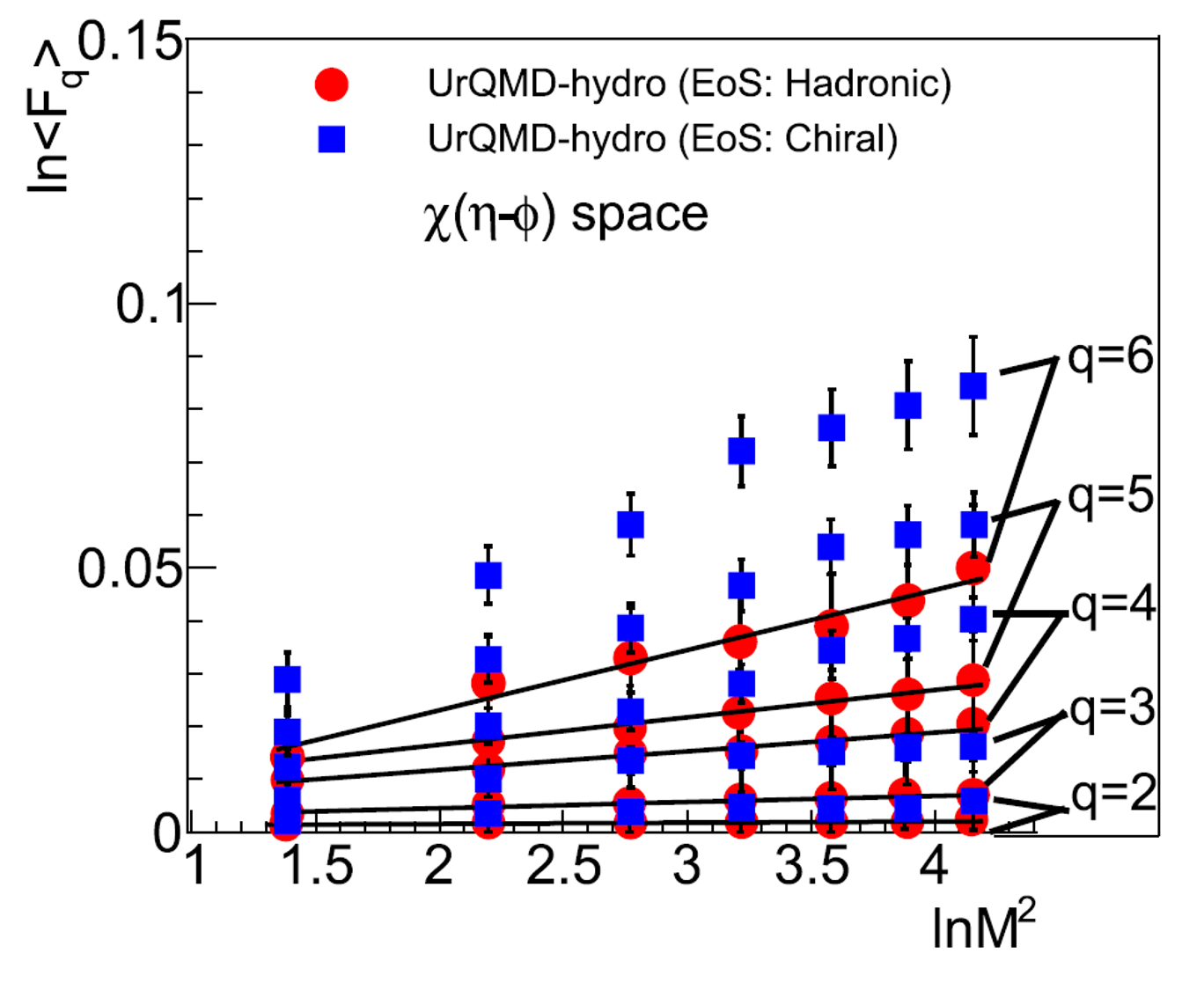}
     \caption{$\ln F_{q}(M)$ vs $\ln M^{2}$ of charged particles in Au+Au collisions at 10A GeV/$c$ from the UrQMD-hydro model with hadronic EoS (red circles) and chiral one (blue squares). Figure taken from Ref~\cite{GopeEPJA}.}
     \label{FigUrQMDhydro}
\end{figure}

In Fig.~\ref{FigUrQMDhydro}, the values of $\ln F_{q}(M)$ are shown as a function of $\ln M^{2}$ of charged particles in Au+Au collisions at 10A GeV/$c$ from the UrQMD-hydro model with two different equations of state (EoS). The $F_{q}(M)$ of various orders are carried out in $\eta-\phi$ space and the background has been subtracted by the cumulative variable method. The value of $\ln F_{q}(M)$ is found to linearly develop with increasing $\ln M^{2}$ for both hadronic EoS (red circles) and chiral one (blue squares) in the (0–5\%) most central collisions. It implies that the observed scaling behavior or intermittency is closely related to the evolution of the medium produced in the collision. In addition, the intermittency indices of chiral EoS are found to be larger than those of hadronic one. It could be account for the production of cascading particles in partonic level and the hydrodynamic evolution in the chiral EoS case~\cite{GopeEPJA}.

\section{Summary and Outlook}
In this review, we have summarized current status of the exploration of the QCD critical point via intermittency analysis in heavy-ion collisions. The main results from both experimental measurements and phenomenological model calculations are presented. We highlighted two important issues, i.e. background subtraction and efficiency correction, in this analysis. It provides a way to obtain a clean signal in the calculations of intermittency.

Although the main contributions from background have been eliminated by using the proposed cumulative variable method, the intermittency index calculated in the transport UrQMD model which does not incorporate any critical self-similar fluctuations is not exact zero, as shown in Fig.~\ref{Fig:F2EUrQMD} and \ref{Fig:F2CentUqMD}. It might attribute to some effects which still remained in the results, such as correlation of protons caused by Coulomb repulsion and Fermi–Dirac statistics~\cite{NA49EPJC}, or width of the experimental momentum resolution~\cite{SamantaJPG2021}. 
More investigations on these effects should help to obtain a clear and convincing result.

On the search of self-similar and intermittency behavior in current heavy-ion experiments, the NA49 Collaboration has found a clear signature in Si+Si collisions at  $\sqrt{s_\mathrm{NN}}$ = 17.3 GeV~\cite{NA49EPJC}, with the measured second-order intermittency index touching the theoretic expectation calculated in a critical model~\cite{AntoniouPRL}. Preliminary results from NA61/SHINE~\cite{NA61CFM,NA61CFM2} of proton numbers in Ar+Sc collisions at 150A GeV/$c$ and in the most central Pb+Pb collisions at 30A GeV/$c$ exhibit no evidence of power-law increase with increasing number of division cells. However, the RHIC/STAR measurement~\cite{STARIntermittency} at the near energy region illustrates a non-monotonic dependence of a scaling exponent on collision energy in the most central Au+Au collisions with a dip at around 20-30 GeV. Further investigations should be done to compare results coming from both the STAR and the NA61 heavy-ion experimental data despite the differences between two detectors.

The RHIC experiment has updated a few detectors and finished taking the second phase of Beam Energy Scan program during 2018-2021~\cite{besII,besII2}. It could be exciting and anticipated if intermittency would be measured by the STAR Collaboration on the new data with high statistics and improved particle identification to explore the CP in the QCD phase diagram.

In addition to the benefits from the upgraded facilities discussed above, the large data samples that collected in the high luminosity phase at BES-II may bring computing difficulties since the calculation of SFMs at very small scales requires a huge computational effort which may prevent its implementation in experimental analysis. A new computational technique, compared to conventional methods, is suggested to be much more efficient and prominent in the calculations of the second-order SFM with increase of the dimension of space~\cite{newMethod}. The technique is also supposed to be a useful tool for recognizing weak signals which may be hidden in the background with strong noises.

With the development of modern computer hardware and artificial intelligence, machine learning (ML)~\cite{ML1Science,ML2Nature}, as a data-driven method, may be another intriguing direction for future investigations. Based on a recent study~\cite{MLIntermittency}, a dynamical edge convolution plus point cloud neural network shows a strong pattern recognition ability in identifying events with self-similar fluctuations. It can figure out the most majority of signal particles which could be used for decision-making in Monte-Carlo events. It is interesting to investigate whether this cutting-edge ML method would be also helpful to pick out possibly existing weak intermittency signals associated with critical phenomena, which encountered in the NA61 experiment.

\section*{Acknowledgments}
I am grateful to many colleagues for conversations that have shaped my understanding of self-similar property and intermittency over the years. In particularly, I would like to express my sincere appreciations to Prof. Lianshou Liu, Yuanfang Wu, Feng liu, Nu Xu, W. Kittel, A. Bialas, W. Metzger, Jinghua Fu, Fuming Liu, Xiaofeng Luo and Lizhu Chen for collaborations and instructive suggestions. I further thank the STAR and the NA61 Collaborations for valuable discussions and comments. This work is supported by the Ministry of Science and Technology (MoST) under grant No. 2016YFE0104800.

\bibliographystyle{ws-mpla}
\bibliography{sample}
\end{document}